\documentstyle[epsfig]{lamuphys}
\begin{document}
\setlength{\unitlength}{\textwidth}
\title
{
Aspects of the Noisy Burgers Equation
}
\author{Hans \, Fogedby}
\institute
{
Institute of Physics and Astronomy,
University of Aarhus, DK-8000, Aarhus C
\and
NORDITA, Blegdamsvej 17, DK-2100, Copenhagen {\O}, Denmark
}
\maketitle
\begin{abstract}
The noisy Burgers equation describing for example the growth
of an interface subject to noise is one of the simplest model
governing an intrinsically nonequilibrium problem.
In one dimension this equation is analyzed by means of the
Martin-Siggia-Rose technique. In a canonical formulation
the morphology and scaling behavior are accessed by a principle
of least action in the weak noise limit. The growth morphology
is characterized by a dilute gas of nonlinear soliton modes
with gapless dispersion law $E\propto p^{3/2}$ and a superposed
gas of diffusive modes with a gap. The scaling exponents and a
heuristic expression for the scaling function follow from a 
spectral representation.
\end{abstract}
\section{Introduction}
Macroscopic phenomena far from equilibrium are ubiquitous and include
phenomena such as turbulence in fluids, interface and growth problems,
chemical reactions, biological processes, and even aspects of economical
and sociological structures.

In recent years much of the focus of modern statistical physics
and soft condensed matter has shifted towards such systems. Drawing on 
the case of static and dynamic critical phenomena in and close to
equilibrium, where scaling, critical exponents, and the concept of
universality have so successfully served to organize our understanding
and to provide a variety of calculational tools, a similar approach
has been advanced towards the much larger class of nonequilibrium
phenomena with the purpose of elucidating scaling properties and more
generally the morphology or pattern formation in a driven state.

In this context the noisy Burgers equation in one dimension provides
maybe the simplest continuum description of an open driven nonlinear
system exhibiting both scaling and pattern formation. This equation has
the form (\cite{Forster76,Forster77})
\begin{eqnarray}
\frac{\partial u}{\partial t} = \nu\nabla^2u + \lambda u\nabla u +\nabla\eta 
\enspace ,
\label{f1}
\end{eqnarray}
and was in the noiseless case for $\eta = 0$ originally proposed
by Burgers (\cite{Burgers74}) in order to model turbulence in fluids;
we note the similarity with the Navier-Stokes equation for $\lambda = -1$.
Equation (\ref{f1}) has the form of a conserved nonlinear
Langevin equation, $\partial u/\partial t = -\nabla j$, with fluctuating
current $j = -\nu\nabla u - (\lambda/2) u^2 + \eta$. The linear diffusive
damping term $\nu\nabla^2 u$ is characterized by the surface tension
or viscosity $\nu$. The nonlinear convective or mode coupling term
$\lambda u\nabla u$ is controlled by $\lambda$. Finally, the equation
is driven by the fluctuating ``white noise'' $\eta$ for which we assume a 
Gaussian amplitude distribution 
\begin{eqnarray}
P(\eta) = \exp{\left[-\frac{1}{2\Delta}\int dxdt \eta(xt)^2\right]}
\enspace ,
\label{f2}
\end{eqnarray}
and short-range correlations  in space according to
the correlation function
\begin{eqnarray}
\langle\eta(xt)\eta(x't')\rangle = \Delta\delta(x-x')\delta(t-t')
\enspace ,
\label{f3}
\end{eqnarray}
characterized by the noise strength parameter $\Delta$.

In the context of modeling a growing interface the Burgers
equation governs the local slope $u=\nabla h$ of a height
field $h$ (in the Monge representation) characterized by the
much studied Kardar-Parisi-Zhang equation 
(\cite{Kardar86,Kardar89})
\begin{eqnarray}
\frac{\partial h}{\partial t} = \nu\nabla^2h+\frac{\lambda}{2}(\nabla h)^2
+ \eta
\enspace .
\label{f4}
\end{eqnarray}
In this case, which we shall adhere to in the following,  $\nu$ is a
diffusion coefficient or viscosity, $\lambda$ a nonlinear
lateral growth parameter, and $\eta$ represents noise in the
drive or the environments. In Fig. 1 we have sketched the growth
morphology in terms of the height and slope fields for a typical
growing interface.
\begin{figure}
\begin{picture}(1.0,0.7)
\put(0.0,-0.4){\epsfig{figure=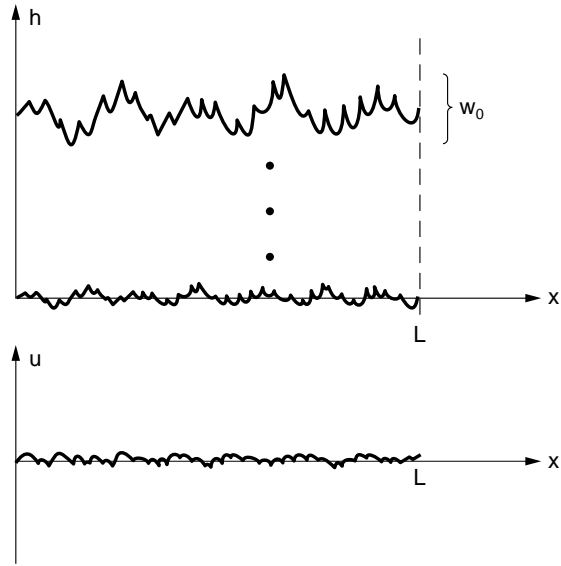,width=1.0\textwidth}}
\end{picture}
\caption[ ]{We depict the growth morphology in terms of the height
($h$) and slope ($u$) fields for a growing interface. The
saturated width in the stationary state is denoted by $w_0$.
}
\end{figure}

The substantial conceptual problems encountered in nonequilibrium\linebreak
physics
are in many ways embodied in the Burgers-KPZ equations (\ref{f1}) and
(\ref{f4})
describing the self-affine growth of an interface subject to
annealed noise arising from fluctuations in the drive or in
the environment.
Interestingly, the Burgers-KPZ equations are also encountered in a variety
of other problems such as randomly stirred fluids,
dissipative transport in a driven lattice gas,
the propagation of flame fronts,
the sine-Gordon equation, and magnetic
flux lines in superconductors. Furthermore, by means of the Cole-
Hopf transformation the Burgers-KPZ equations are also
related to
the problem of a directed polymer or
a quantum particle in a random
medium  and thus to the theory of spin glasses; see e.g.
(\cite{Halpin95}).

One issue which has been addressed is the scaling properties
of the noisy Burgers equation. In Fig. 2 we have depicted how the
width of the fluctuating interface (see Fig. 1) after a transient
lapse of time, characterized by the a crossover time $t_{crossover}$
scaling with
the system size $L$ according to $t_{crossover}\propto L^z$, 
where $z$ is the dynamic 
exponent, saturates to a value $w_0\propto L^\zeta$ also depending
on $L$ and characterized by the roughness exponent $\zeta$.

\begin{figure}
\begin{picture}(1.0,0.4)
\put(0.0,-0.8){\epsfig{figure=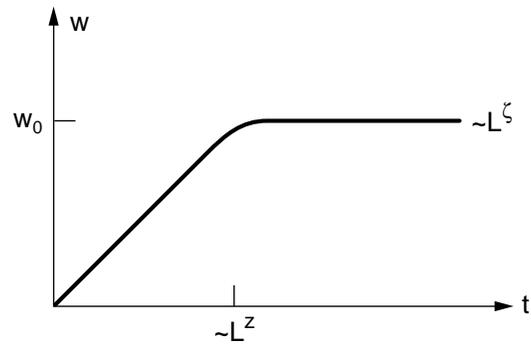,width=1.2\textwidth}}
\end{picture}
\caption[ ]{We depict the interface width $w(t)$ as a function of
time $t$. In the transient regime $t\ll t_{crossover}\sim L^{\zeta}$
$w$ grows according to $t^{\zeta/z}$. In the stationary regime attained for
$t\gg t_{crossover}$ $w$ saturates to the value $w_0\sim L^{\zeta}$.
}
\end{figure}

This dynamical scaling hypothesis, which is substantiated by numerical
modeling and renormalization group considerations, is embodied in the 
following relationship for the width:
\begin{eqnarray}
w(t,L)=L^\zeta F_1(t/L^z)
\enspace .
\label{f5}
\end{eqnarray}
In terms of the height field itself or the associated slope field
the appropriate dynamical scaling forms are given by
\begin{eqnarray}
\langle[h(x+x_0,t+t_0)-h(x_0,t_0)]^2\rangle = x^{2\zeta}F_2(t/x^z) 
\enspace ,
\label{f6}
\end{eqnarray}
and
\begin{eqnarray}
\langle u(x+x_0,t+t_0)u(x_0,t_0)\rangle = x^{2\zeta-2}F_3(t/x^z)
\enspace .
\label{f7}
\end{eqnarray}
Here $x_0$ and $t_0$ are reference points in an infinite system
in the stationary regime and $F_i$ the scaling functions.
The scaling issue thus amounts to a determination of the scaling
exponents $\zeta$ and $z$ together with the scaling functions $F_i$.

The hydrodynamical origin of the Burgers equation, 
as reflected by the presence of the mode coupling or convective
term $\lambda u\nabla u$, 
implies that the Burgers equation is invariant under a Galilean 
transformation
\begin{eqnarray}
x&&\rightarrow x-\lambda u_0t 
\label{f8}
\\
u&&\rightarrow u+u_0 
\enspace ,
\label{f9}
\end{eqnarray}
involving a shift of the slope field. Since the nonlinear
coupling strength $\lambda$ enters as a structural constant
in the symmetry group it is invariant under scaling. 
This property implies a dynamical scaling law
\begin{eqnarray}
\zeta + z = 2
\enspace ,
\label{f10}
\end{eqnarray}
relating the roughness and dynamic exponents.

Another interesting property of the Burgers equation specific
to one dimension is the existence of an effective static
fluctuation-dissipation theorem, in the sense that the stationary
Fokker-Planck equation for the Burgers equation is solved by
a Gaussian distribution
\begin{eqnarray}
P_{\mbox{st}}(u)\propto\exp{\left[-\frac{\nu}{\Delta}\int dx u(x)^2\right]}
\label{f11}
\end{eqnarray}
for the slope field {\em independent} of the nonlinear coupling 
strength $\lambda$.
This implies that the slope field $u$ performs independent 
Gaussian fluctuations and that, consequently, the height field
$h=\int udx$, performs random walk in $x$, yielding the roughness
exponent and from the scaling law (\ref{f10}) also the dynamic
exponent. Summarizing, the universality class for the Burgers
equation is characterized by the exponents:
\begin{eqnarray}
\zeta = \frac{1}{2}~~~~~~~ \mbox{and}~~~~~~~z=\frac{3}{2}
\enspace .
\label{f12}
\end{eqnarray}
By a combination of the Galilean invariance and the effective
fluctuation-dissipation theorem specific to one dimension the
exponents in one dimension are thus exactly known. The scaling function
has also been accessed numerically and by a mode coupling approach.
In higher dimensions the scaling properties of the Burgers-KPZ
are controversial and remain to be clarified. In this context
the dynamic renormalization group based on an expansion in 
$\lambda^2\Delta/\nu^3$ and an expansion about the 
critical dimension $d=2$
yields limited results and in particular fails to access the
strong coupling fixed point in $d=1$, characterized by the
exponents in (\ref{f12}).

In a recent series of papers 
(\cite{Fogedby95,Fogedby98})
we have presented a novel approach
to the strong coupling features of the noisy Burgers equation
in one dimension based on a nonperturbative approach in the
asymptotic weak noise limit. Since the singular character of the
weak noise limit is an essential feature of our approach and is 
already apparent in the linear case for $\lambda = 0$ it is
instructive first to consider this case.
\section{The Edwards -- Wilkinson Equation}
In the linear case for $\lambda = 0$ the Burgers equation
(\ref{f1}) takes the form of the Edwards-Wilkinson (EW)
equation (for the slope field) (\cite{Edwards82})
\begin{eqnarray}
\partial u/\partial t = \nu\nabla^2u + \nabla\eta ~ ,
\label{f13}
\end{eqnarray}
i.e., a linear diffusion equation driven by conserved noise.

Equation (\ref{f13}) is readily analyzed. The time-dependent
probability distribution for the wavenumber modes
$u_k = \int dx u(x)\exp{(-ikx)}$, $u_k^{\ast}=u_{-k}$, is given by
\begin{eqnarray}
P(\{u_k\},t|\{u_k^0\})\propto
\prod_k\exp{\left[-\frac{\nu}{\Delta}\frac{1}{L}
\frac{|u_k - u_k^0 e^{-\nu k^2t}|^2}
{1-e^{-2\nu k^2t}}\right]}
\enspace ,
\label{f14}
\end{eqnarray}
where $u_k^0$ is the initial value for $t=0$. For $t\rightarrow\infty$
$P(\{u_k\},t)$ approaches the stationary distribution  (\ref{f11}),
using $(1/L)\sum_k|u_k|^2 = \int dx u(x)^2$.

In a similar way we obtain for the slope correlations in $\omega k$-space,
$u(k\omega) = \int dxdt\exp{(i\omega t-ikx)}$, the Lorentzian line
shape characteristic of a conserved diffusive mode,
\begin{eqnarray}
\langle u(k\omega)u(-k-\omega)\rangle = \frac{\Delta k^2}{\omega^2 + 
(\nu k^2)^2}
\enspace ,
\label{f15}
\end{eqnarray}
implying the static correlations $\langle u(x)u(x')\rangle =
(\Delta/2\nu)\delta(x-x')$ in conformity with (\ref{f11}).
In Fig. 3 we have depicted the correlation function.
\begin{figure}
\begin{picture}(1.0,0.5)
\put(0.0,-0.5){\epsfig{figure=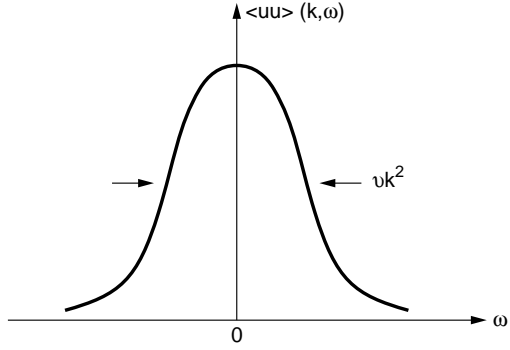,width=1.0\textwidth}}
\end{picture}
\caption[ ]{We depict the slope correlation function for the diffusive
mode in the EW case. The Lorentzian is centered about $\omega = 0$ and has
the hydrodynamical linewidth $\nu k^2$ vanishing in the long
wave length limit.
}
\end{figure}

Comparing  (\ref{f15}) with the scaling form in  (\ref{f7})
we also infer the scaling exponents $\zeta =1/2$ and $z=2$, 
characteristic of diffusion and defining the EW universality class.
Also, noting that the diffusive term in  (\ref{f13}) can be
derived from a free energy $F = (1/2)\int dx u^2$, it follows that
the EW equation  describes the fluctuations in an equilibrium
system with ``temperature'' $\Delta/2\nu$, i.e.,
$P_{\mbox{st}} = \exp{[-(2\nu/\Delta)F]}$.

We note already here that the noise strength $\Delta$ plays a special
role. Whereas $\Delta$ enters linearly in the correlation function
(\ref{f15}), the limit $\Delta\rightarrow 0$ appears as an 
``essential singularity'' in the distribution functions
$P(\{u_k\},t)$ and $P_{\mbox{st}}(u)$ in  (\ref{f11}) and (\ref{f14}).
The special role of the noise as the relevant small parameter is
more physically recognized by considering the time-dependent correlations
\begin{eqnarray}
&&\langle u(kt)u(-kt')\rangle
=\nonumber
\\
&&[\langle uu\rangle_i
- (\Delta/2\nu)]\exp{[-(t+t')\nu k^2]}
+(\Delta/2\nu)\exp{[-|t-t'|\nu k^2]} 
\enspace ,
\label{f16}
\end{eqnarray}
$\langle\cdots\rangle_i$ denotes an initial value average.
The basic time scale is set by $\tau(k) = 1/\nu k^2$ which diverges
for $k\rightarrow 0$, characteristic of a hydrodynamical mode.
However, there is another ``time scale'' set by the noise, namely
a characteristic crossover time
\begin{eqnarray}
\tau(\Delta)\sim\tau(k)\ln{\frac{1}{\Delta}}
\enspace .
\label{f17}
\end{eqnarray}
For $t,t'\ll\tau(k)$ the correlations are nonstationary and depend
on the initial correlation $\langle\cdots\rangle_i$, whereas for
long times $t,t'\gg\tau(k)$ the correlations enter a stationary,
time reversal invariant regime and depends on $|t-t'|$.
The crossover time $\tau(\Delta)$ defines the onset of the
stationary regime. For $t,t'\gg\tau(k),\tau(\Delta)$ noise-induced
fluctuations built up and the system becomes stationary; for
$\Delta\rightarrow 0$, $\tau(\Delta)\rightarrow\infty$, and the system
never leaves the transient regime, i.e., there is no stationary regime.

Whereas the singular nature of the weak noise limit is easily understood
in the context of the linear EW equation; it is in fact precisely
equivalent to the low temperature limit of the Boltzmann factor
$\exp{(-F/T)}, T=\Delta/2\nu$, the situation is more subtle in the
Burgers equation where the nonlinear growth term leads to a general 
nonequilibrium system.

In order to understand the role of the nonlinear term we first consider
the well-understood properties of the noiseless or deterministic
Burgers equation.
\section{The Noiseless Burgers Equation}
The noiseless Burgers equation has the form
\begin{eqnarray}
\partial u/\partial t = \nu\nabla^2u + \lambda u\nabla u
\enspace ,
\label{f18}
\end{eqnarray}
and describes the transient decay of an interface subject to the
damping term $\nu\nabla^2u$ in combination with the mode coupling 
or nonlinear growth term $\lambda u\nabla u$. Correspondingly,
for the height field we obtain the deterministic KPZ equation
\begin{eqnarray}
\frac{\partial h}{\partial t} = \nu\nabla^2h+\frac{\lambda}{2}(\nabla h)^2
\enspace .
\label{f19}
\end{eqnarray}
Interestingly, the noiseless Burgers equation (\ref{f18}) can be 
solved analytically by means of the nonlinear Cole-Hopf transformation
(\cite{Cole51} and \cite{Hopf50})
\begin{eqnarray}
w =&& \exp{\left[\frac{\lambda}{2\nu}\int dx u\right]}
\enspace ,
\label{f20}
\end{eqnarray}
which maps (\ref{f18}) onto the linear diffusion equation
\begin{eqnarray}
\frac{\partial w}{\partial t} = \nu\nabla^2w
\enspace ,
\label{f21}
\end{eqnarray}
and an initial value analysis can be carried out, in particular
in the inviscid limit $\nu\rightarrow 0$. The basic transient
mode structure consists of i) solitons or shock waves, connected
by ii) ramp modes, and iii) superposed diffusive modes.
\subsection{Linear Modes for $\lambda = 0$}
In the linear case for $\lambda = 0$ the noiseless Burgers equation
reduces to the ordinary diffusion equation
\begin{eqnarray}
\frac{\partial u}{\partial t} = \nu\nabla^2 u
\enspace ,
\label{f22}
\end{eqnarray}
which supports decaying diffusive modes
\begin{eqnarray}
u(xt) \propto \exp{[-i\omega_k^0t + ikx]}
\enspace ,
\label{f23}
\end{eqnarray}
with gapless quadratic dispersion	
\begin{eqnarray}
\omega_k^0 = -i\nu k^2
\enspace .
\label{f24}
\end{eqnarray}
In Fig. 4 we have depicted the diffusive spectrum

\begin{figure}
\begin{picture}(1.0,0.5)
\put(0.0,-0.55){\epsfig{figure=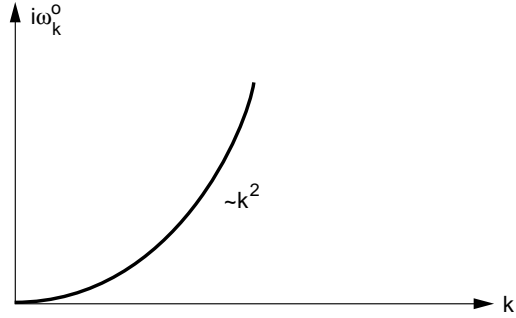,width=1.0\textwidth}}
\end{picture}
\caption[ ]{
We show the gapless dispersion law for the diffusive modes
for the EW case.
}
\end{figure}
\subsection{Soliton and Ramp Modes for $\lambda\neq 0$}
In the presence of the nonlinear mode coupling term the
noiseless Burgers equation, violating parity symmetry,
supports a localized ``right hand'' 
soliton mode of the sine-Gordon form
\begin{eqnarray}
u_k(x) = u_+\tanh{k_s(x-x_0)}
\enspace ,
\label{f25}
\end{eqnarray}
with center position $x_0$ and amplitude $u_+$. The inverse width
\begin{eqnarray}
k_s = \frac{\lambda u_+}{2\nu}
\enspace ,
\label{f26}
\end{eqnarray}
depends on the amplitude; in the inviscid limit $\nu\rightarrow 0$,
$k_s\rightarrow\infty$, and the soliton  becomes a sharp shock wave.
Owing to the Galilean invariance in (\ref{f8}-\ref{f9}) we
also easily obtain a boosted soliton moving with velocity $v$.
Denoting the boundary values at $x=\pm\infty$ by $u_\pm$, we infer
the single soliton condition
\begin{eqnarray}
u_+ +u_- = -\frac{2v}{\lambda}
\enspace ,
\label{f27}
\end{eqnarray}
expressing the soliton velocity in terms of the boundary values
$u_\pm$.
The height profile corresponding to a soliton has the form
\begin{eqnarray}
h(xt) = \frac{2\nu}{\lambda}\ln{\cosh{[k_s(x-x_0)]}}
\enspace .
\label{f28}
\end{eqnarray}
The soliton mode is a stable ``dissipative structure''corresponding to
deterministic input at the boundaries
(nonvanishing currents) and dissipation at the soliton center (hot spot).
In Fig. 5 we have depicted a moving soliton and the
associated height profile.
\begin{figure}
\begin{picture}(1.0,0.8)
\put(0.0,-0.4){\epsfig{figure=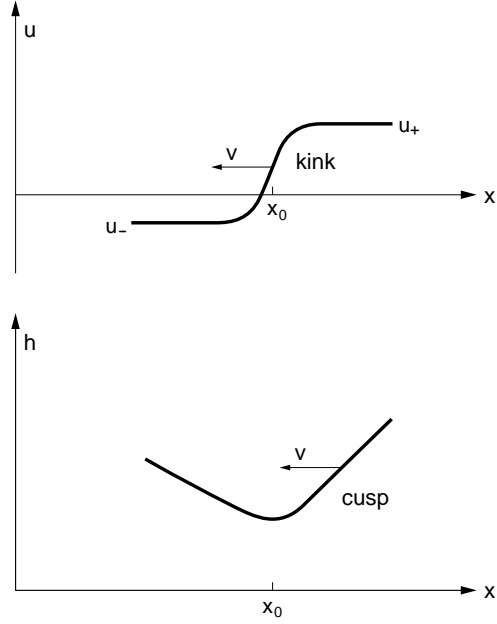,width=1.0\textwidth}}
\end{picture}
\caption[ ]{
We depict a moving ``right hand'' soliton for the '
noiseless Burgers equation together
with the associated downward cusp in the height profile.
}
\end{figure}

Furthermore, in the inviscid limit $\nu\rightarrow 0$ (\ref{f18}) supports
a negative-slope ramp solution
\begin{eqnarray}
u(xt)\propto -\frac{x}{\lambda t}
\enspace ,
\label{f29}
\end{eqnarray}
corresponding to a convex parabolic height profile
\begin{eqnarray}
h(xt)\propto -\frac{x^2}{2\lambda t}
\enspace .
\label{f30}
\end{eqnarray} 
In Fig. 6 we have shown the ramp solution and the associated
height profile.
\begin{figure}
\begin{picture}(1.0,0.8)
\put(0.0,-0.4){\epsfig{figure=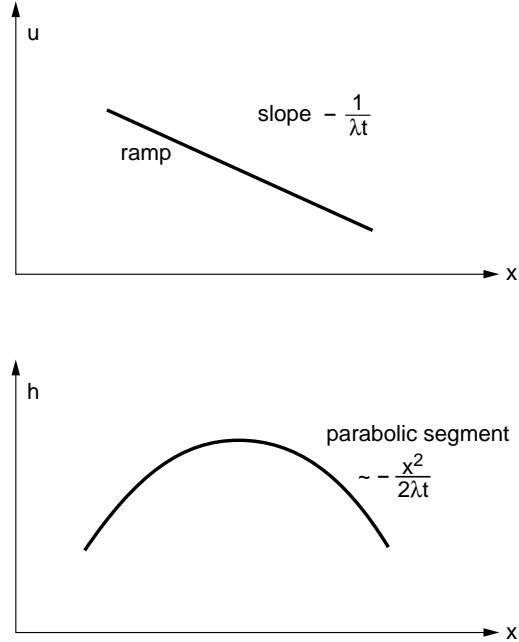,width=1.0\textwidth}}
\end{picture}
\caption[ ]{
We show a ramp solution for the noiseless Burgers equation together
with the associated parabolic height profile.
}
\end{figure}
\subsection{Superposed Linear Modes}
The character of the diffusive modes in the presence of a soliton
is examined by means of a linear stability analysis of Eq. (\ref{f18}).
Inserting
\begin{eqnarray}
u = u_k + \delta u
\enspace ,
\label{f31}
\end{eqnarray}
and solving the associated linear eigenvalue problem we obtain,
in addition to a localized translation mode, a band of
damped diffusive modes
\begin{eqnarray} 
\delta u\propto \exp{[-i\omega_kt + ikx]}
\enspace ,
\label{f32}
\end{eqnarray}
with a soliton-induced gap in the dispersion law
\begin{eqnarray}
\omega_k = -i\nu(k^2 + k_s^2), ~~~~~~ k_s = \frac{\lambda u_+}{2\nu}
\enspace .
\label{f33}
\end{eqnarray} 
In Fig. 7 we have shown the diffusive spectrum in the presence of
the soliton.
\begin{figure}
\begin{picture}(1.0,0.5)
\put(0.0,-0.65){\epsfig{figure=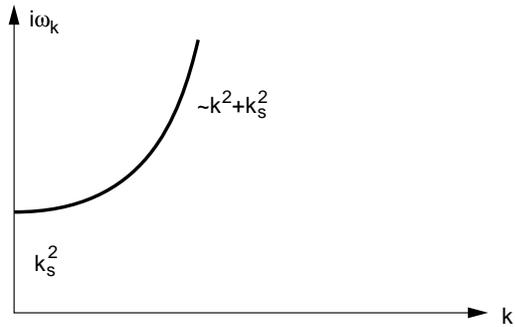,width=1.0\textwidth}}
\end{picture}
\caption[ ]{
We depict the diffusive dispersion law in the presence of a soliton
for the noiseless Burgers equation. The gap is proportional to
the soliton amplitude.
}
\end{figure}

The soliton modes and associated superposed diffusive modes are thus
an essential feature of the nonlinear noiseless Burgers equation;
this aspect will 
persist in the noisy case.
The morphology of the noiseless Burgers equation is clear: The
equation is basically damped but in the transient regime dissipative
structures are present consisting of an initial value induced
gas of ``right hand'' solitons connected by ramps with superposed linear
diffusive modes. With vanishing slope field at the boundaries
there is no deterministic current-input and the morphology
eventually decays through interaction and coalescence of moving
solitons.
In Fig. 8 we have shown the transient morphology of the noiseless
Burgers equation and the associated height profile.
\begin{figure}
\begin{picture}(1.0,0.4)
\put(0.0,-0.55){\epsfig{figure=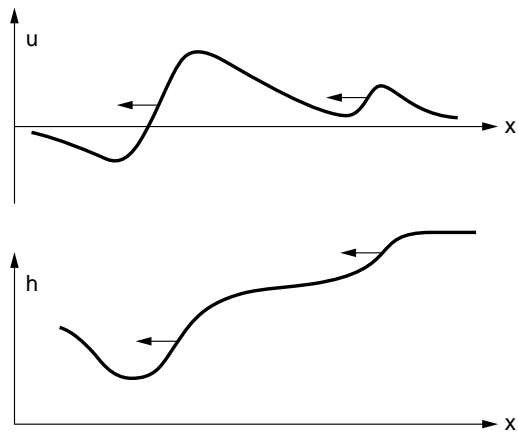,width=1.0\textwidth}}
\end{picture}
\caption[ ]{
We show the transient morphology for the noiseless Burgers equation
consisting of ``right hand'' solitons connected by ramps for
the slope field $u$ and downward cusps connected by parabolic
segments for the associated height field $h$.
}
\end{figure}
\section{The Noisy Burgers Equation}
In Sects. 2 and 3 we discussed the linear EW case in order
to illustrate the singular character of the weak noise limit
and the noiseless Burgers case in order to identify
the essential nonlinear feature, i.e., the soliton or shock wave, in 
determining the morphology.
Here we perform an analysis of the noisy Burgers equation (\ref{f1})
including both the nonlinear soliton modes and taking into account the 
singular character of the noise.
\subsection{Phase Space Path Integral Formulation}
The Martin-Siggia-Rose technique (\cite{Martin73}) in a path 
integral formulation (\cite{Zinn89})
is the appropriate language to analyze the Burgers equation and works in
the following way:
All the information about the statistical properties of the Burgers
equation is basically contained in the generator or characteristic
function
\begin{eqnarray}
Z(\mu) = \langle\exp{\left[i\int dxdt\mu u\right]}\rangle
\enspace .
\label{f34}
\end{eqnarray}
The probability distribution $P(\{u(x)\},t)$ and correlation
functions are easily extracted from $Z$, e.g., 
\begin{eqnarray}
\langle u(xt)u(x't')\rangle = 
-[\delta Z/\delta\mu(xt)\delta\mu(x't')]_{\mu=0}
\enspace .
\label{f35}
\end{eqnarray}
Here $\langle\cdots\rangle$ denotes an average withe respect to the noise
$\eta$. In order to implement the Burgers equation (\ref{f1}) which provides
the statistical link between the output slope field and the input noise
driving the field, we insert the identity
$\int\prod_{xt}duJ\delta(\partial u/\partial t-\nu\nabla^2 u - 
\lambda u\nabla u-\nabla\eta) = 1$ in (\ref{f34})(\cite{Footnote}), expand the 
delta function, $\delta(x)\propto\int dp\exp{(ipx)}$, and average
over the noise according to the Gaussian distribution in  (\ref{f2}).
This procedure yields the path integral
\begin{eqnarray}
Z(\mu)\propto\int\prod_{xt}dudp\exp{\left[\frac{i}{\Delta}\tilde{S}\right]}
\exp{\left[i\int dxdt\mu u\right]}
\enspace ,
\label{f36}
\end{eqnarray}
where the action $\tilde{S}$ has the form
\begin{eqnarray} 
\tilde{S} = \int dxdt\left[p\frac{\partial u}{\partial t} - {\cal H}\right]
\enspace ,
\label{f37}
\end{eqnarray}
with Hamiltonian
\begin{eqnarray}
{\cal H} = -\frac{i}{2}(\nabla p)^2 + p(\nu\nabla^2u+\lambda u\nabla u)
\enspace .
\label{f38} 
\end{eqnarray}
In the linear case for $\lambda = 0$ we expect the Hamiltonian to reduce
to a harmonic form. This is accomplished by performing a shift operation
of the noise variable $p$. Setting $p = \nu(iu - \varphi)$ we obtain
the canonical phase space form
\begin{eqnarray}
Z(\mu)\propto\int\prod_{xt}dud\varphi\exp{\left[i\frac{\nu}{\Delta}S\right]}
\exp{\left[i\frac{\nu}{\Delta}S_B\right]} \exp{\left[i\int dxdt\mu u\right]}
\enspace ,
\label{f39}
\end{eqnarray}
with bulk action
\begin{eqnarray}
S = \int dxdt\left[u\frac{\partial\varphi}{\partial t} - {\cal H}\right]
\enspace ,
\label{f40} 
\end{eqnarray} 
and canonical Hamiltonian
\begin{eqnarray} 
{\cal H} = -i\frac{\nu}{2}[(\nabla u)^2 + (\nabla\varphi)^2]
+ \frac{\lambda}{2}u^2\nabla\varphi
\enspace .
\label{f41}  
\end{eqnarray}  
The shift transformation also generates a surface term
$S_B$ which contributes to the action for a particular
slope configuration, depending on the choice of boundary condition. 
It will not play any role in the present discussion; see (\cite{Fogedby98}). 

The Hamiltonian ${\cal H}$ is composed of a relaxational or irreversible
harmonic component characterizing the EW case, and a nonlinear
reversible mode coupling component associated with the drive $\lambda$
in the Burgers-KPZ case.

The path integral (\ref{f39}) has the usual Feynman form in  phase space
with $u$ and $\varphi$ as canonically conjugate variables, $u$ is the
canonical ``momentum'' and the noise variable $\varphi$ the canonical
``coordinate''. We note, however, that $\cal H$ does not break up in 
a ``kinetic energy'' part and a ``potential energy'' part as is 
characteristic of an ordinary dynamical problem; the coordinate and 
momentum variables are mixed in the nonlinear interaction term.
Furthermore, the Hamiltonian is complex characteristic of the 
Master equation for a relaxational
problem.

Nevertheless, the Feynman form is very suggestive and allows us to draw
both on ``classical mechanics'' in analyzing the path integral and
on the associated ``quantum field theory,'' yielding the path integral.
We note that the noise strength $\Delta/\nu$ plays the role of an 
effective Planck constant defining a ``correspondence principle''
associated with the weak noise limit $\Delta\rightarrow 0$.

There are essentially two ways to approach the path integral (\ref{f39}):
i) renormalized perturbation theory in the nonlinear term $\lambda$ and
ii) a nonperturbative saddle point calculation for small $\Delta$.

Perturbation theory in $\lambda$ yields the same expansion that is 
obtained by directly expanding the noisy Burgers equation (\ref{f1})
and contracting the noise term by term according to (\ref{f3}).
This expansion which is logarithmically divergent in the infrared limit
in $d=2$ and algebraically divergent below $d=2$ requires a 
regularization scheme below $d=2$ yielding renormalization group 
equations for $\nu$, $\lambda$, and $\Delta$. While $\lambda$ is 
invariant under scaling (yielding the scaling law in (\ref{f10})),
the effective expansion coefficient is $\lambda^2\Delta/\nu^3$
yielding the exponents $(\zeta,z) = (1/2,3/2)$ in $d=1$. Note that
the expansion is effectively in $\Delta$ and thus does not retrieve
singular terms in $\Delta$.

As discussed earlier the noise $\Delta$ is the small parameter in the
problem, at least in $d=1$ where we have carried out the analysis.
The singular terms in $\Delta$ are precisely obtained by performing a
saddle point analysis of the path integral, corresponding to the
WKB or ``quasi-classical'' limit of the Feynman path integral.
\subsection{Field Equation for $\Delta\rightarrow 0$}
In the weak noise limit $\Delta\rightarrow 0$ the path integral
formulation allows for ``a principle of least action'' in that the
leading saddle point or stationary contributions are obtained from
the variational principle $\delta S = 0$ with respect to variations
$\delta u$ and $\delta\varphi$. We thus obtain the Hamiltonian
equations of motion
\begin{eqnarray}
\frac{\partial u}{\partial t} = -\frac{\delta H}{\delta\varphi}
~~~~~~~~\mbox{and} ~~~~~~~~~
\frac{\partial\varphi}{\partial t} = +\frac{\delta H}{\delta u}
\enspace ,
\label{f42}
\end{eqnarray}
where $H = \int dx{\cal H}$ and inserting (\ref{f41})
the field equations
\begin{eqnarray}
\frac{\partial u}{\partial t}&&
=-i\nu\nabla^2\varphi + \lambda u\nabla u
\label{f43}
\\
\frac{\partial\varphi}{\partial t} &&
=+i\nu\nabla^2 u  + \lambda u\nabla \varphi 
\enspace .
\label{f44}
\end{eqnarray}
The coupled deterministic field equations (\ref{f43}-\ref{f44}) are an
essential result of our analysis. In the weak noise limit 
$\Delta\rightarrow 0$ they effectively replace the noisy Burgers
equation (\ref{f1}).

Note that we here have obtained a definite level of simplification in
handling the statistical problem. The 
stochastic character of the Langevin equation (\ref{f1}) has been replaced
by the path integral, yielding the additional variable $\varphi$
characterizing the noise, and in the weak noise limit  the coupled
field equations (\ref{f43}-\ref{f44}). The alternative formulation
in terms of the deterministic Fokker-Planck equation 
is technically more
difficult since the Fokker-Planck equation is a functional-differential
equation, in fact ``the Schr\"{o}dinger equation'' for the
path integral.
In Fig. 9 we have depicted the paths in $u\varphi$ phase space.
\begin{figure}
\begin{picture}(1.0,0.5)
\put(0.0,-0.5){\epsfig{figure=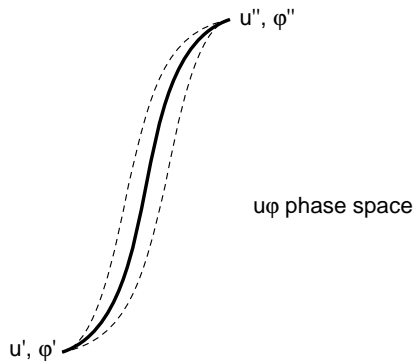,width=1.0\textwidth}}
\end{picture}
\caption[ ]{
We depict the paths in $u\varphi$ phase space connecting the
initial values $u',\varphi'$ to the final values $u'',\varphi''$.
The fully drawn extremal path corresponds to the stationary or saddle point
solution for $\Delta\rightarrow 0$; the dashed paths characterize the 
fluctuations about the stationary solution.
}
\end{figure}
\subsection{Linear Diffusive Modes}
It is instructive first to discuss the linear EW case. For $\lambda = 0$
the field equations (\ref{f43}-\ref{f44}) reduce to the linear coupled
pair
\begin{eqnarray} 
\frac{\partial u}{\partial t}&& 
=-i\nu\nabla^2\varphi
\label{f45} 
\\ 
\frac{\partial\varphi}{\partial t} && 
=+i\nu\nabla^2 u 
\enspace ,
\label{f46} 
\end{eqnarray} 
admitting the solution
\begin{eqnarray}
u(xt) \sim u^+_0\exp{[-i\omega_k^0t + ikx]}
+ u^-_0\exp{[i\omega_k^0t - ikx]}
\label{f47}
\end{eqnarray}
with gapless dispersion given by (\ref{f24}). We note the doubling of solutions,
i.e., the equations imply both a damped and a growing solution, unlike
the noiseless Burgers equation discussed in Sect. 3.1. This is a 
feature of the noisy case: In the stationary regime we attain time
reversal invariance and both solutions are required in order to describe
the stationary correlations.

For $\lambda = 0$ the path integral is Gaussian and we leave it as an 
exercise to derive the correlation function (\ref{f15}); it is also
not difficult to derive the distribution (\ref{f14}).
\subsection{Nonlinear Soliton Modes}
Leaving aside the question whether the field equations (\ref{f43}-\ref{f44})
in the nonlinear case for $\lambda\neq 0$ are exactly integrable or
admit a Cole-Hopf type transformation, we find, like in the case of the
noiseless Burgers equation, permanent profile soliton
solutions, see e.g.  (\cite{Fogedby81}). In the static case we have
\begin{eqnarray}
u_0(x) =\pm u_+\tanh{k_s(x-x_0)}
\label{f48}
\end{eqnarray}
of the same form as (\ref{f25}). Note, however, again a doubling of the
solutions like in the linear case. In the stationary regime the
noise excites 
both ``right hand'' and ``left hand'' solitons. Both  modes are 
required in order to correctly describe the stationary growth morphology.
In Fig.10 we have depicted the two soliton solutions and the associated 
height field.
\begin{figure}
\begin{picture}(1.0,0.6)
\put(-0.1,-0.15){\epsfig{figure=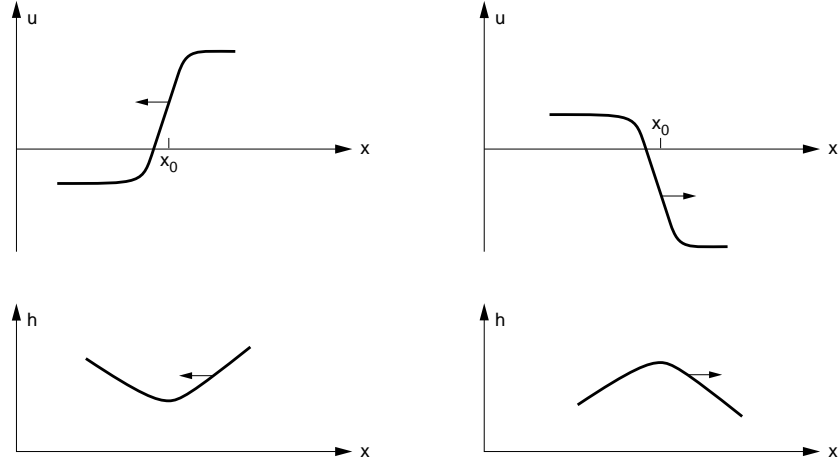,width=1.2\textwidth}}
\end{picture}
\caption[ ]{
We depict the  ``right hand'' and ''left hand'' solitons
for the noisy Burgers equation and the associated upward and
downward cusps in the associated height field.
}
\end{figure}
\subsection{Superposed Linear Modes}
Like in the noiseless case we can perform a linear stability analysis
of the field equations by inserting $u=u_0+\delta u$ and 
$\varphi= \varphi_0 +\delta\varphi$. The resulting linear eigenvalue 
problem is exactly soluble, see e.g. (\cite{Fogedby85}). 
In addition to a translation mode associated
with a displacement of the soliton position $x_0$, we obtain a band
of modes
\begin{eqnarray}
\delta u \sim u^+\exp{[-i\omega_kt + ikx]}
+ u^-\exp{[i\omega_kt - ikx]}
\label{f49}
\end{eqnarray}
with a diffusive dispersion law with a gap given by Eq. (\ref{f33}).
In Fig. 11 we have depicted a ``right hand'' soliton with a superposed
diffusive mode.
\begin{figure}
\begin{picture}(1.0,0.4)
\put(0.0,-0.6){\epsfig{figure=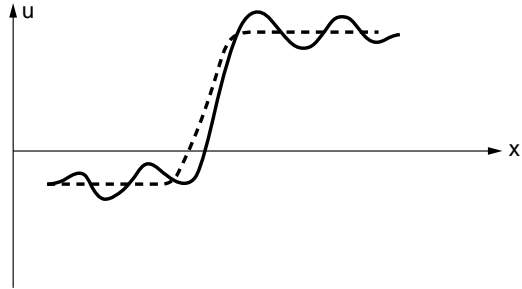,width=1.0\textwidth}}
\end{picture}
\caption[ ]{
We depict a ''right hand'' soliton (the dashed curve) 
with a superposed diffusive mode (the solid curve); the 
superposed damped linear mode is phase-shifted and exhibits a
gap in the spectrum.
}
\end{figure}
\setlength{\unitlength}{\textwidth}
\begin{figure}
\begin{picture}(1.0,0.7)
\put(0.0,-0.4){\epsfig{figure=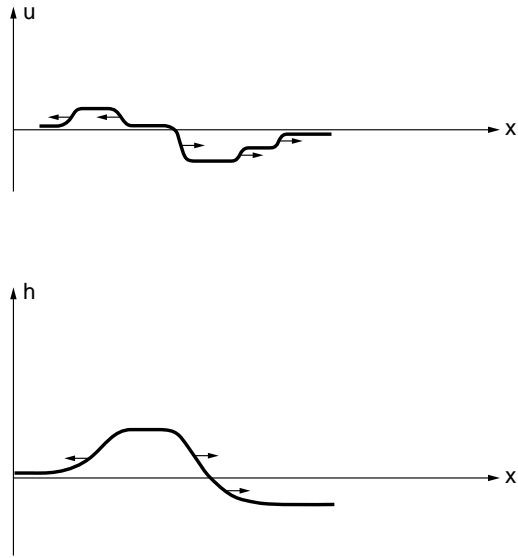,width=1.0\textwidth}}
\end{picture}
\caption[ ]{
We show the stationary growth morphology for the noisy Burgers equation,
consisting of ``right hand'' and ``left hand'' solitons connected
by horizontal segments for the slope field $u$ and downward and
upward cusps connected by constant slope segments for the
associated height field $h$.
}
\end{figure}
\subsection{A Growing Interface as a Dilute Soliton Gas}
The saddle point modes, i.e., the solitons and diffusive modes,
excited by the noise immediately allow a physical description
of a growing interface. First we observe that the saddle point 
approximation actually implies a dilute gas of solitons connected
by constant slope segments and satisfying the soliton condition
(\ref{f27}); in the limit of well-separated solitons these
configurations also provide saddle point solutions (\cite{Zinn89}).
A single soliton corresponds to a downward or upward cusp in $h$.
A pair of solitons describe a moving step, etc. In Fig. 12 we have 
depicted a general growth morphology in the slope field $u$
and the associated height field $h$. Superposed on the soliton
configuration is a gas of diffusive modes
(not indicated on the figure). We note that the stationary growth 
morphology is quite different from the transient morphology
shown in Fig. 8.
\subsection{Dynamics of Solitons}
The Hamiltonian structure of the path integral allows for 
``a principle of least action'' and we can associate dynamic
and kinetic attributes to the soliton and diffusive modes.
The energy is inferred from (\ref{f41}),
\begin{eqnarray}
E = \int dx\left[-i\frac{\nu}{2}[(\nabla u)^2 + (\nabla\varphi)^2]
+ \frac{\lambda}{2}u^2\nabla\varphi\right]
\enspace .
\label{f50}
\end{eqnarray}
The momentum, i.e., the generator of translation, has a form inferred
from the Poisson bracket $\{u(x),\varphi(x')\}=\delta(x-x')$
for the canonically conjugate variable $u$ and $\varphi$,
\begin{eqnarray}
P = \int dx u\nabla\varphi
\enspace .
\label{f51}
\end{eqnarray} 
For a single soliton we have in terms of the boundary values
\begin{eqnarray} 
E =&& \pm i\frac{\lambda}{6}[u_+^3 - u_-^3]
\label{f52}
\\
P =&& \pm i\frac{1}{2}[u_+^2 - u_-^2]
\enspace ,
\label{f53} 
\end{eqnarray}  
where $\pm$ indicates the ``right hand'' and ``left hand''
solitons, respectively.

In particular, for a two-soliton configuration representing
a growing step with vanishing slope at the boundaries
we obtain
\begin{eqnarray} 
E =&& \frac{8}{3}i\frac{|v|^3}{\lambda^2}
\label{f54} 
\\
P =&& -4i\frac{v^2}{\lambda^2}sgn(v)
\enspace .
\label{f55}  
\end{eqnarray}   
We note the nonlinear velocity dependence characteristic of
a soliton mode, see e.g. (\cite{Fogedby81}). 
Eliminating the velocity we infer the
soliton dispersion law
\begin{eqnarray}
E = i\lambda\frac{\sqrt{2}}{3}|P|^{3/2}
\enspace ,
\label{f56}   
\end{eqnarray}   
and we recover the dynamic exponent $z=3/2$ characterizing
the gapless soliton dispersion. In Fig. 13 we have depicted the
soliton dispersion law.
\begin{figure}
\begin{picture}(1.0,0.5)
\put(0.0,-0.6){\epsfig{figure=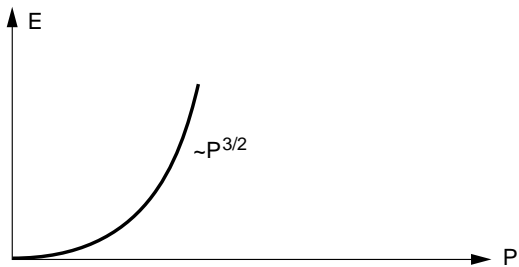,width=1.0\textwidth}}
\end{picture}
\caption[ ]{
We show the gapless soliton dispersion law characterized
by the ``fractional'' exponent $3/2$.
}
\end{figure}
\subsection{Stationary Distribution}
As alluded to in Sect. 1 it is known that the stationary 
distribution in the Burgers-KPZ case is Gaussian and given by
(\ref{f11}) (\cite{Huse85}). Here we attempt to make this result, 
which follows
easily from the Fokker-Planck equation, plausible within the
Martin-Siggia-Rose formalism.

From  (\ref{f36}) it follows that the stationary distribution
is given by
\begin{eqnarray} 
P_{\mbox{st}}(u'')\propto\lim_{T\rightarrow 0}\int_{u'}^{u''}
\prod_{xt}dpdu\exp{\left[\frac{i}{\Delta}\tilde{S}\right]}
\enspace ,
\label{f57}    
\end{eqnarray}    
where $u'=u(x,0)$ and $u''=u(x,T)$ are the initial and final values
of the slope field and $\tilde{S}$ is integrated from $t=0$ to
$t=T$. In the weak noise limit $\Delta\rightarrow 0$ variation
of $\tilde{S}$ yields  saddle point equations equivalent to
(\ref{f43}-\ref{f44}), 
\begin{eqnarray} 
\frac{\partial u}{\partial t} =&& \nu\nabla^2u+\lambda u\nabla u+i\nabla^2p
\label{f58}
\\
\frac{\partial p}{\partial t} =&& -\nu\nabla^2p+\lambda u\nabla p
\enspace .
\label{f59} 
\end{eqnarray}     
Using that the energy vanishes in the stationary state attained in the
limit $T\rightarrow\infty$, we obtain from (\ref{f37}) at
the saddle point
\begin{eqnarray}
P_{\mbox{st}}(u)\propto\lim_{T\rightarrow\infty}\left[\frac{i}{\Delta}
\int^Tdxdtp\frac{\partial u}{\partial t}\right]
\enspace ,
\label{f60}
\end{eqnarray}     
where $p$ and $u$ are solutions of the field equations (\ref{f58}-\ref{f59}).
In order to demonstrate that $-ip\rightarrow 2\nu u$ for
$T\rightarrow\infty$ implying the stationary distribution (\ref{f11}),
$P_{\mbox{st}}\propto\exp{[-(\nu/\Delta)\int dx u^2]}$, we define the 
deviation $\Delta u$ according to $-ip=2\nu(u+\Delta u)$ and find to 
linear order
\begin{eqnarray}
\left[\frac{\partial}{\partial t}-\lambda u\nabla\right]\Delta u = 
\nu\nabla^2\Delta u
\enspace .
\label{f61} 
\end{eqnarray}      
Owing to the Galilean invariance of the operator 
$\partial/\partial t-\lambda u\nabla$ we can choose an instantaneous frame
with vanishing  $u$ and (\ref{f61}) implies a decaying solution
$\Delta u\propto\exp{(-\nu k^2 t)}$. This is not a rigorous proof;
basically we assume that the trajectories on the $E=0$ energy surface
in $u\varphi$ phase space at long times are attracted to the 
sub manifold defined by $-ip=2\nu u$.
\section{Scaling and Universality Classes}
The present path integral approach also permits a simple discussion
of the scaling properties of the noisy Burgers equation.
Focussing on the slope correlation function (\ref{f39})
yields
\begin{eqnarray}
\langle u(xt)u(x't')\rangle\propto\int\prod dud\varphi
\exp{[i(\nu/\Delta)S]}u(xt)u(x't')
\enspace ,
\label{f62}
\end{eqnarray}
where we assuming vanishing boundary conditions have set
$S_B=0$. The direct evaluation of the path integral (\ref{f62}) requires 
the application of methods from quantum chaos such as periodic orbit 
theory, see e.g. (\cite{Dashen74}) and is still in progress.
However, by discussing $\langle uu\rangle$ in terms of the underlying
``quantum field theory'' it is an easy task to extract the scaling
properties.

In the ``quantum description'' the canonical fields $u$ and $\varphi$
are replaced by ``quantum operators'' $\hat{u}$ and $\hat{\varphi}$
and likewise the Hamiltonian and momentum in (\ref{f41}) and (\ref{f51}).
The correlation function (\ref{f62}) can thus be expressed as the 
time-ordered product (\cite{Zinn89})
\begin{eqnarray}
\langle u(xt)u(00)\rangle\propto
\langle 0|T\hat{u}(xt)\hat{u}(00)|0\rangle
\enspace ,
\label{f63}
\end{eqnarray}
where $\hat{u}$ evolves in time according to the ``quantum Hamiltonian''
(\ref{f41}) and $|0\rangle$ denotes the zero-energy stationary state.
Displacing the field from $(x,t)$ to $(0,0)$, using the Hamiltonian and
momentum operators, we have
\begin{eqnarray}
\hat{u}(xt) = \exp{[i(\nu/\Delta)(\hat{P}x+\hat{H}t)]}
\hat{u}(00)
\exp{[-i(\nu/\Delta)(\hat{P}x+\hat{H}t)]}
\enspace ,
\label{f64}
\end{eqnarray}
which inserted in (\ref{f63}) together with a complete set
of intermediate quasi-particle wavenumber states $|K\rangle$,
$P=(\Delta/\nu)K$, with frequency $\Omega$, $E=(\Delta/\nu)\Omega$,
yields the spectral representation
\begin{eqnarray}
\langle u(xt)u(00)\rangle\propto\int dKG(K)\exp{[-i(\Omega t-Kx)]}
\enspace .
\label{f65}
\end{eqnarray}
Here $G(K)$ is a form factor and $\Omega ,K$ are the frequencies and
wavenumbers of the quasi-particles in the theory.

The scaling limit for large $x$ and large $t$ corresponds to the
bottom of the quasi-particle spectrum and we note that only gapless
excitations contribute. Assuming a general dispersion law with 
exponent $\beta$,
\begin{eqnarray}
\Omega=AK^\beta
\label{f66}
\enspace ,
\end{eqnarray}
and assuming that the form factor $G(K)$ is regular for small wavenumber,
$G(K)\propto\mbox{const.}$, we obtain, rescaling $K$
\begin{eqnarray}
\langle u(xt)u(00)\rangle\propto x^{-1}\int dK
e^{-iAK^\beta(t/x^\beta)-iK}
\label{f67}
\enspace .
\end{eqnarray}
Comparing the spectral form (\ref{f67}) with the dynamic scaling
hypothesis (\ref{f7}) we first infer the robust roughness exponent 
$\zeta=1/2$, independent of the quasi-particle dispersion law.
The dynamic exponent $z$ is given by the exponent $\beta$ for the
quasi-particle dispersion law. In the linear EW case the gapless diffusive
dispersion law $\Omega\propto k^2$ yields the dynamic exponent
$z=2$; in the Burgers-KPZ case the diffusive modes develop a gap
and do not contribute to the scaling; however, the noise excites 
a new nonlinear gapless soliton mode with dispersion $\Omega\propto K^{3/2}$,
yielding the exponent $z=3/2$. We also obtain a heuristic expression
for the scaling function $F_3$ in (\ref{f7}),
\begin{eqnarray}
F_3(w)\propto\int dKe^{-i(K^zw+K)}
\enspace ,
\label{f69}
\end{eqnarray}
which has the same form as the probability distribution for 
L\'{e}vy flights (\cite{Fogedby94}).
In Table 1 we have summarized the exponents and universality classes
for the EW and Burgers-KPZ cases.
\begin{table}
\caption[]{Scaling exponents and universality classes}
\renewcommand{\arraystretch}{2.2}
\begin{tabular}{lccl}
\hline
Model&Roughness exp. ($\zeta$)&~~~~~Dynamic exp. ($z$)~~~~~&Universality class\\
\hline
EW equation       &  $\frac{1}{2}$ &    $2$            & EW\\
\hline
Burgers equation  &  $\frac{1}{2}$ &    $\frac{3}{2}$  & Burgers-KPZ\\
\hline
\end{tabular}
\end{table}
\section{Summary and conclusion}
We have here advanced a novel approach to the growth
morphology and scaling behavior of the noisy Burgers equation
in one dimension. Using the Martin-Siggia-Rose (MSR) technique in a canonical
form we have demonstrated that the physics of the
strong coupling
fixed point  is associated with an essential singularity
in the noise strength and can be accessed by appropriate theoretical
soliton techniques.

The canonical representation of the MSR functional integral in terms
of a Feynman phase space path integral with a complex Hamiltonian
identifies the noise strength as the relevant small nonperturbative
parameter and allows for {\em a principle of least action}. In the
asymptotic weak noise limit the leading contributions to the path
integral are given by a dilute gas of solitons with superposed
linear diffusive modes. The canonical variables are the local slope
of the interface and an associated ``conjugate'' noise field, characteristic
of the MSR formalism. In terms of the local slope the soliton and diffusive
mode picture provides a many-body description of a growing interface governed
by the noisy Burgers equation. The noise-induced slope fluctuations
are here represented by the various paths or configurations contributing
to the path integral. Moreover, a spectral representation of the slope
correlations based on the underlying ``quantum field theory'' gives
access to the scaling exponents and provide a heuristic expression
for the scaling function.

So far the present nonperturbative approach has only been implemented
in the one dimensional case where the analysis is tractable.
However, the singular nature of the weak noise limit seems to be
a general feature of the onset of the stationary regime and might 
also be important for the Burgers-KPZ equations in higher dimension.
It remains to be investigated whether the present saddle point
approach can be generalized to this case.

\vspace{.5cm}
\noindent
{\bf Acknowledgments}

\vspace{.2cm}
\noindent
Discussions with J. Krug, M. Kosterlitz, M. H. Jensen, T. Bohr, M. Howard,
K. B. Lauritsen and A. Svane
are gratefully acknowledged.

%
% ---- Bibliography ----
%


\begin{thebibliography}
\bibitem{}{Footnote}{see note on the Jacobian $J$}
The Jacobian $J = |\partial(\partial u/\partial t)/
\partial(u)|$ only contributes to the fluctuations
about the saddle point(s) and will not enter in the
present discussion
\end{thebibliography}
\end{document}